\documentclass[twocolumn,tighten]{aastex62}

\usepackage{amsmath}
\usepackage{xspace}
\usepackage{multirow}
\usepackage{fancyapj}
\usepackage{lipsum}

\newcommand{\E}[1]{\ensuremath{\times 10^{#1}} }

\newcommand{\msol}{\ensuremath{M_{\odot}}\xspace}
\newcommand{\rxte}{\textit{RXTE}\xspace}
\newcommand{\xmm}{\textit{XMM-Newton}\xspace}

\newcommand{\nicer}{\textit{NICER}\xspace}
\newcommand{\low}{\ensuremath{\ell}ow\xspace}

\newcommand{\aql}{Aql~X-1\xspace}

\begin{document}

\title{A NICER look at the Aql X-1 hard state}

\author{Peter Bult}
\affiliation{Astrophysics Science Division, 
  NASA's Goddard Space Flight Center, Greenbelt, MD 20771, USA}
  
\author{Zaven Arzoumanian}
\affiliation{Astrophysics Science Division, 
  NASA's Goddard Space Flight Center, Greenbelt, MD 20771, USA}
 
\author{Edward M. Cackett}
\affiliation{Department of Physics \& Astronomy, Wayne State University, 
  666 W. Hancock, MI 48201, USA}
  
\author{Deepto Chakrabarty}
\affil{MIT Kavli Institute for Astrophysics and Space Research, 
  Massachusetts Institute of Technology, Cambridge, MA 02139, USA}
  
\author{Keith C. Gendreau}
\affiliation{Astrophysics Science Division, 
  NASA's Goddard Space Flight Center, Greenbelt, MD 20771, USA}

\author{Sebastien Guillot} 
\affil{CNRS, IRAP, 9 avenue du Colonel Roche, BP
  44346, F-31028 Toulouse Cedex 4, France} 
\affil{Universit\'e de Toulouse, CNES, UPS-OMP, F-31028 Toulouse, France}
  
\author{Jeroen Homan}
\affil{Eureka Scientific, Inc., 2452 Delmer Street, Oakland, CA 94602, USA}
\affil{SRON, Netherlands Institute for Space Research,
    Sorbonnelaan 2, 3584 CA Utrecht, The Netherlands}
  
\author{Gaurava K. Jaisawal}
\affil{National Space Institute, Technical University of Denmark, 
  Elektrovej 327-328, DK-2800 Lyngby, Denmark}

\author{Laurens Keek}
\affil{Department of Astronomy, University of Maryland, 
  College Park, MD 20742, USA}   
  
\author{Steve Kenyon}
\affil{Astrophysics Science Division, 
  NASA's Goddard Space Flight Center, Greenbelt, MD 20771, USA}

\author{Frederick K. Lamb}
\affil{Center for Theoretical Astrophysics and Department of Physics,
  University of Illinois at Urbana-Champaign,
  1110 West Green Street, Urbana, IL 61801-3080, USA}
\affil{Department of Astronomy, 
  University of Illinois at Urbana-Champaign, 
  1002 West Green Street, Urbana, IL 61801-3074, USA}
  
\author{Renee Ludlam}
\affil{Department of Astronomy, University of Michigan, 
  1085 South University Avenue, Ann Arbor, MI 48109-1107, USA}
 
\author{Simin Mahmoodifar} 
\affil{Astrophysics Science Division and Joint Space-Science Institute, 
  NASA's Goddard Space Flight Center, Greenbelt, MD 20771, USA}
\affil{CRESST II and the Department of Astronomy, 
  University of Maryland, College Park, MD 20742, USA}

\author{Craig Markwardt}
\affil{Astrophysics Science Division, 
  NASA's Goddard Space Flight Center, Greenbelt, MD 20771, USA}
  
\author{Jon M. Miller}
\affil{Department of Astronomy, University of Michigan, 
  1085 South University Avenue, Ann Arbor, MI 48109-1107, USA}
  
\author{Gregory Prigozhin}
\affil{MIT Kavli Institute for Astrophysics and Space Research, 
  Massachusetts Institute of Technology, Cambridge, MA 02139, USA}

\author{Yang Soong}
\affil{Astrophysics Science Division, 
  NASA's Goddard Space Flight Center, Greenbelt, MD 20771, USA}
\affil{Department of Astronomy, University of Maryland, 
  College Park, MD 20742, USA}   

\author{Tod E. Strohmayer} 
\affil{Astrophysics Science Division and Joint Space-Science Institute,
  NASA's Goddard Space Flight Center, Greenbelt, MD 20771, USA}

\author{Phil Uttley}
\affil{Anton Pannekoek Institute, University of Amsterdam,
  Postbus 94249, 1090 GE Amsterdam, The Netherlands}

\begin{abstract} 
  We report on a spectral-timing analysis of the neutron star low-mass
  X-ray binary Aql~X-1  with the \textit{Neutron Star Interior Composition
  Explorer} (NICER) on the International Space Station. Aql~X-1 was 
  observed with \textit{NICER}
  during a dim outburst in 2017 July, collecting approximately $50$ ks 
  of good exposure. The spectral and timing properties of the source 
  correspond to that of an (hard) extreme island state in the atoll 
  classification. 
  We find that the fractional amplitude of the low frequency ($<0.3$ Hz) 
  band-limited noise shows a dramatic turnover as a function of energy: 
  it peaks at 0.5 keV with nearly 25\% rms, drops to $12\%$ rms at 2 keV, 
  and rises to $15\%$ rms at 10 keV. 
  Through the analysis of covariance spectra, we demonstrate 
  that band-limited noise exists in both the soft thermal emission
  and the power-law emission. Additionally, we measure hard time lags, 
  indicating the thermal emission at $0.5$ keV leads the power-law emission
  at 10 keV on a timescale of $\sim100$ ms at $0.3$ Hz to $\sim10$ ms 
  at $3$ Hz. Our results demonstrate that the thermal emission in the 
  hard state is intrinsically variable, and
  driving the modulation of the higher energy power-law. Interpreting the 
  thermal spectrum as disk emission, we find our results are consistent
  with the disk propagation model proposed for accretion onto black holes.
\end{abstract}


\section{Introduction} \label{sec:intro}
    The X-ray transient Aquila X-1 is a neutron star low-mass X-ray
    binary (LMXB) in a 19 h orbit with a $\sim1\msol$ companion star
    \citep{Thorstensen1978,Mata2017}. The system shows frequent outbursts
    with a recurrence rate that evolved from $\sim125$ days
    \citep{Priedhorksy1984} to $\sim300$ days \citep{Kitamoto1993},
    to the approximate $\sim250$ days in recent times
    \citep{Campana2013}. These outbursts show a
    remarkable variation in shape and peak luminosity, with dim
    outbursts limited to $\leq 5\%~L_{\rm edd}$ and bright outbursts
    exceeding $30\%~L_{\rm edd}$, for an Eddington luminosity of
    $L_{\rm edd} = 3.8\E{38} \mbox{~erg s}^{-1}$ \citep{Kuulkers2003,
    Campana2013}, and a neutron star mass of $1.4$ Solar mass.

    Over the course of an outburst \aql moves through a series of 
    accretion states, each with distinct spectral and timing
    properties, based on which it has been 
    classified as an `atoll' type neutron 
    star binary \citep{Hasinger1989,Reig2000}. At low luminosities
    the source is in an (extreme) island state, where its energy spectrum 
    is dominated by hard power-law emission and the power spectrum is 
    characterized by broad, large amplitude components. At higher luminosity 
    the source transitions to the so-called `banana' branch, where the energy
    spectrum pivots to be dominated by its soft thermal 
    components \citep[e.g.][]{Lin2007}. Meanwhile
    the variability decreases in amplitude as it shifts to higher frequencies.
    
    Qualitatively similar systematic behavior may be observed in black hole
    LMXBs \citep{KleinWolt2008}, suggesting that the spectral 
    and timing properties must be a consequence of the accretion process, and
    be independent of the type of central object. 
    This idea is reinforced by the relation between the characteristic 
    frequencies of the power spectrum, which scale across source types
    \citep{Wijnands1999}. Because the variability amplitudes tend to 
    increase as a function of energy \citep{Sobolewska2006}, it is thought
    the power-law emitting region modulates the observed X-ray flux, although
    for neutron stars a boundary layer may also play an important role
    \citep{Gilfanov2003}, and in either case the physical process that sets the
    frequency may still originate elsewhere in the accretion system.
    
    Advances in spectral-timing analysis of black hole systems has shown that
    even in the hard state, the soft component may show large amplitude variations
    at slow time scales \citep{Wilkinson2009}, and that this variability leads the
    modulation of the hard power-law \citep{Uttley2011}. These observations may 
    be explained by an intrinsically variable accretion disk that 
    propagates variability down to the power-law emitting region by 
    modulating the mass accretion rate \citep{Lyubarskii1997,Kotov2001,Arevalo2006}. 
    
    It is important to confirm whether the spectral-timing characteristics of the
    band-limited noise are independent of source type, or if they are unique to black
    hole systems. This type of analysis, however, has been difficult to extend to 
    neutron stars, which are generally fainter in the hard state than their black 
    hole counterparts. 
    
    The Neutron Star Interior Composition Explorer (\nicer;
    \citealt{Gendreau2017}) is an external attached X-ray telescope payload
    on the International Space Station (ISS). It combines good spectral 
    and timing resolution with excellent sensitivity at 1 keV. These 
    properties make \nicer a great platform for spectral-timing studies.
    
    In this letter we report on \nicer observations of \aql during
    a dim outburst in 2017 July. We present a spectral-timing analysis of
    all collected data, placing a focus on the behavior of the low
    frequency band-limited noise at low energies.

\section{Observations}
    \nicer's X-ray Timing Instrument (XTI; \citealt{Gendreau2016})
    provides an array of 56 co-aligned X-ray concentrator optics, each
    paired with a silicon drift detector \citep{Prigozhin2012}. \aql
    was observed with 52 detectors in operation, as four detectors
    showed malfunctions prior to launch. Operating in the $0.2-12$ keV
    energy band with a resolution of $\sim100$ eV, these detectors 
    provide a combined $\sim1900$ cm$^2$ 
    effective area at 1.5 keV.  Detected photon events are time-tagged 
    at a relative time resolution of $\sim40$ ns, and have an absolute 
    timing accuracy of $\sim100$ ns rms.

    \nicer observed \aql between 2017 June 20 (MJD 57924) and 2017 
    July 3 (MJD 57937), {collecting $70$ ks of unfiltered exposure.}
    In this letter we consider all available data {(ObsIDs 0050340101
    through 0050340109)}. 
    We process the data using \textsc{heasoft} version 6.22.1 and
    \textsc{nicerdas} version 2018-02-22\_V002d, {selecting only those
    epochs collected with a pointing offset smaller than $54\arcsec$, more than
    $40\arcdeg$ away from the bright Earth limb, more than $30\arcdeg$ away
    from the dark Earth limb, and outside of the South Atlantic Anomaly (SAA)}. 
    Because the first four ObsIDs were collected while calibration of the X-ray
    boresight was ongoing, these data have a pointing offset of 70\arcsec.
    {Since these four ObsIDs contain less than 5\% of the available
    exposure, we did not attempt to account for the off-axis instrument
    response, and instead exclude them from the present analysis.} 
    
    We then construct a light curve using a $12-15$ keV energy range. This range
    lies above the nominal energy band of the instrument because above
    12 keV the performance of the optics and detectors has diminished 
    such that essentially no astronomical signal is expected.
	In the last ObsID, at MJD 57937.545, we observed a 200 second epoch 
    where the $12-15$ keV light curve is significantly different from 
    zero. Correlated with this, the $0.4-10$ keV count-rate also increases, 
    which we attribute to a high-background interval. This {200-s} 
    epoch is therefore excluded from our analysis.
    Finally we use the \textsc{ftool} \textsc{barycorr} with the radio 
    position of \citet{Tudose2013} to adjust the photon arrival times 
    to the Solar System barycenter (DE405). 

    After processing we are left with 51 ks of good exposure.  Due
    to the low-Earth orbit of the ISS, our observations are frequently
    interrupted by Earth occultations and passages through the SAA. 
    Typical continuous exposures are between 100 and 1000 seconds in length. 
    
    The source count-rate was $\sim5$ counts/s/detector,
    yielding a total rate of 270 counts/s. {Because \nicer does
    not provide imaging capabilities, we estimated the background rate
    from \nicer observations of the \rxte background field 5
    \citep{Jahoda2006}. We estimate the background} contributed 1.5
    counts/s in the $0.4-10$ keV band we consider in this work.
    
    Two type I X-ray bursts were observed by \nicer, the initial
    analysis of which has been reported by \citet{Keek2018}. We
    excluded the X-ray burst epochs from the analysis presented
    here, where we defined each epoch as the 200 s interval beginning
    $50$ s prior to the burst start time. 

\section{Timing}
    For the stochastic timing analysis we constructed a light curve at
    a 1/8192-s time resolution. We divided this light curve into 
    16-s segments and, for each segment, computed the  power 
    spectrum as well as a soft ($1.1-2.0$ keV/$0.5-1.1$ keV) and hard 
    ($3.8-6.8$ keV/$2.0-3.8$ keV) color ratio. Because neither the colors
    nor the light curve count-rate showed significant evolution over 
    the course of the observations, we averaged all segments to a 
    single Leahy-normalized power spectrum \citep{Leahy1983a}.
    
    Visual inspection of the high frequency ($\geq 2000$ Hz) powers
    confirmed that the power distribution converges to a mean value of
    $2$, consistent with Poisson statistics \citep{Leahy1983a}. This is
    in line with the expectation that deadtime effects are not
    significant at the recorded count-rates due to the modularity of
    \nicer's design. We therefore proceeded by subtracting a constant
    power level of 2 from our spectrum and renormalized the powers to
    obtain fractional rms amplitudes with respect to the source rate 
    \citep{Klis1995}. 

    The final power spectrum is modeled using a multi-Lorentzian 
    component model \citep{Belloni2002}, with each Lorentzian 
    $L(\nu ; r, Q, \nu_{\rm max})$ defined by the characteristic 
    frequency $\nu_{\rm max} = \nu_0 \sqrt{1 + 1/4Q^2}$, quality 
    factor $Q$, and centroid frequency $\nu_0$. The rms amplitude, 
    $r$, is defined as 
    \begin{equation}
        r^2 = \int_0^{\infty} L(\nu) d\nu.
    \end{equation}

\begin{figure}
    \centering
    \includegraphics[width=\linewidth]{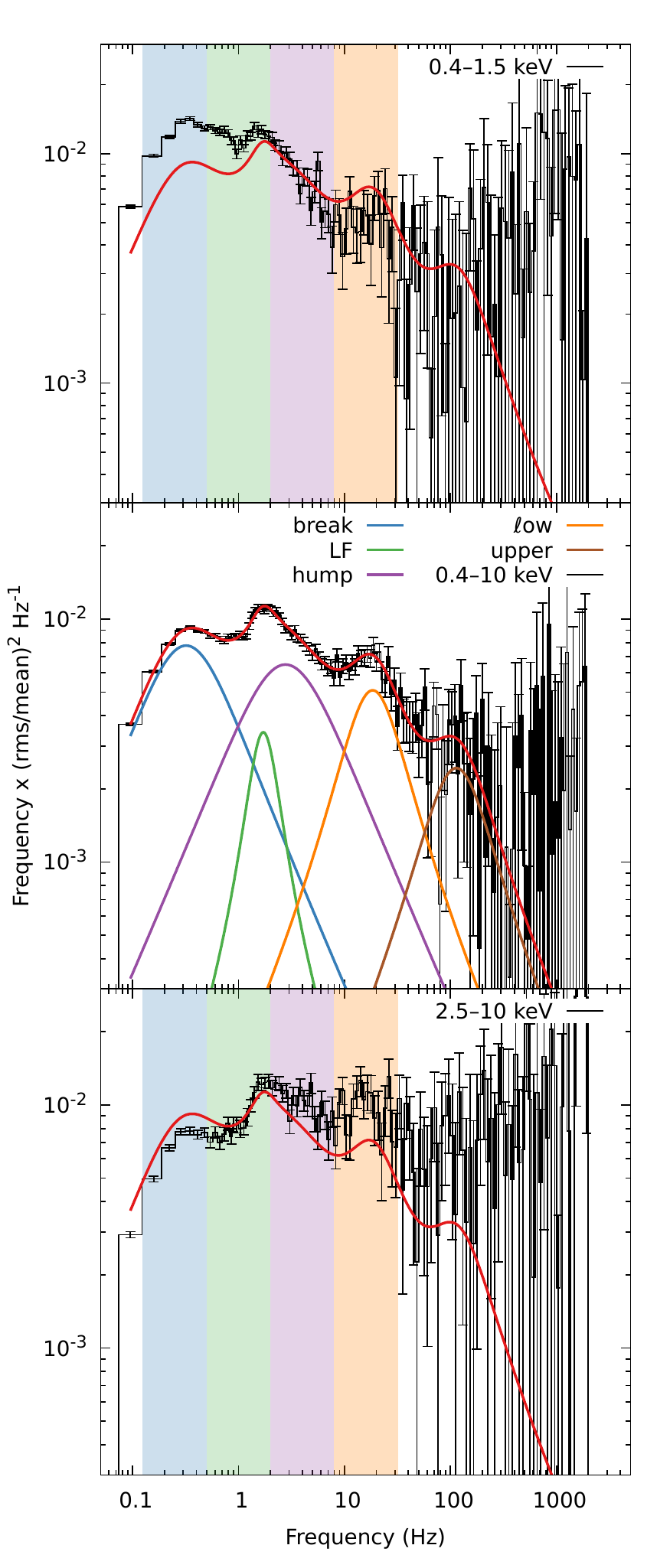}
    \caption{%
        \aql power spectra and the best-fit $0.4-10$ keV model (red), 
        with; top to bottom: the $0.4-1.5$ keV, $0.4-10$ keV, 
        and $2.5-10$ keV band data. {Top and bottom panels are
        included for illustrative purposes only.} Shaded regions indicate
        the covariance frequency intervals (see section \ref{sec:st}).
    }
    \label{fig:pds}
\end{figure}

    The power spectrum of \aql is well described by a five-component 
    model (see Figure \ref{fig:pds}), the best-fit parameters of 
    which are shown in Table \ref{tab:fit}. 
 	The morphology of the power spectrum 
    is that of the extreme island state \citep{Reig2004}, allowing us 
    to identify the components as (for increasing frequency): the
    break, low-frequency (LF), hump, \low\footnote{%
    	The \low component, traditionally formatted as such, is distinct
        from the lower kHz quasi-periodic oscillation \citep[see, e.g,][]{Klis2006}.
    }, and upper kHz {(see \citealt{Klis2006} for a description
    of neutron star timing nomenclature)}. We note that none
    of the components are sufficiently coherent to qualify as a 
    quasi-periodic oscillation (the threshold is usually set at 
    $Q = 2$).

\begin{table}[t]
    \newcommand{\mc}[1]{\multicolumn2c{#1}}
    \centering
	\caption{%
    	Power spectrum fit parameters
		\label{tab:fit}	
	}
    \begin{tabular}{l D D D c }
    \decimals
	\tableline
    ~     & \mc{Frequency} &\mc{Quality}& \mc{Fractional} & $\chi^2$ / dof \\
    ~     & ~              &\mc{factor} & \mc{amplitude}  &  \\
    ~     & \mc{(Hz)}      &\mc{~}      & \mc{(\% rms)}   &  \\
	\tableline
    break &  0.320(10)  &  0.21(2)   &  14.2(0.4)   & \multirow{5}{*}{115/116} \\
    LF    &  1.71(4)    &  1.3(0.3)  &   6.0(1.4)   & \\
    hump  &  2.8(0.4)   &  0.14(8)   &  13.5(1.0)   & \\
    \low  &  18.4(0.8)  &  0.57(14)  &   9.6(0.9)   & \\
    upper &  114. (19)  &  0.5(0.4)  &   6.9(1.1)   & \\
    \tableline
	\end{tabular}
    \flushleft
    \tablecomments{Values in parentheses indicate $1\sigma$ uncertainties.}
\end{table}

    We also searched for lower frequency components by considering the
    averaged power spectrum of 1024 s segments. However, no such
    slow variability is detected, with the power spectrum remaining
    roughly constant below the break frequency of 0.3 Hz. 

    To determine the energy dependence of the power spectrum
    components we divide the 0.4--10 keV energy band into 25 bins, with
    the energy boundaries chosen such that each bin contains roughly
    the same number of counts. We then construct an average power
    spectrum for each bin and fit our multi-Lorentzian model. As
    the frequency and quality factor do not change significantly with energy
    we keep these parameters fixed, allowing only the rms amplitude 
    of each component to vary.

    We find that the amplitude of each component tends to increase
    approximately linearly with energy above 2 keV, consistent with
    previous findings for \aql \citep{Cui1998}. Below 2 keV, however, we see a
    sharp rise in the amplitude of the break component, with $15\%$ 
    rms at 10 keV, $12\%$ rms at 2 keV and a maximum amplitude of 
    $24.5\%$ rms at 0.5 keV.

\section{Spectral-Timing}
\label{sec:st}
    To better characterize the joint spectral and temporal variations,
    we compute time lags and covariance spectra \citep{Uttley2014}
    for four frequency bands between $0.125$ Hz and $32$ Hz. We scale these 
    frequency ranges geometrically, such that each band is roughly a factor
    $4$ larger than the previous one. These bands correspond approximately
    with the break, LF, hump and \low components of the power spectrum. A higher 
    frequency band corresponding to the upper kHz term was left out, as it contained
    insufficient power to meaningfully constrain the covariance. 
    Spectral-timing products are calculated for narrow energy bands
    {($\sim200$ eV)} with respect to a broad $0.5-10$ keV reference band 
    by cross correlating each narrow band with the sum of all other bands 
    \citep{Wilkinson2009}.
    
\begin{figure}[t]
    \centering
    \includegraphics[width=\linewidth]{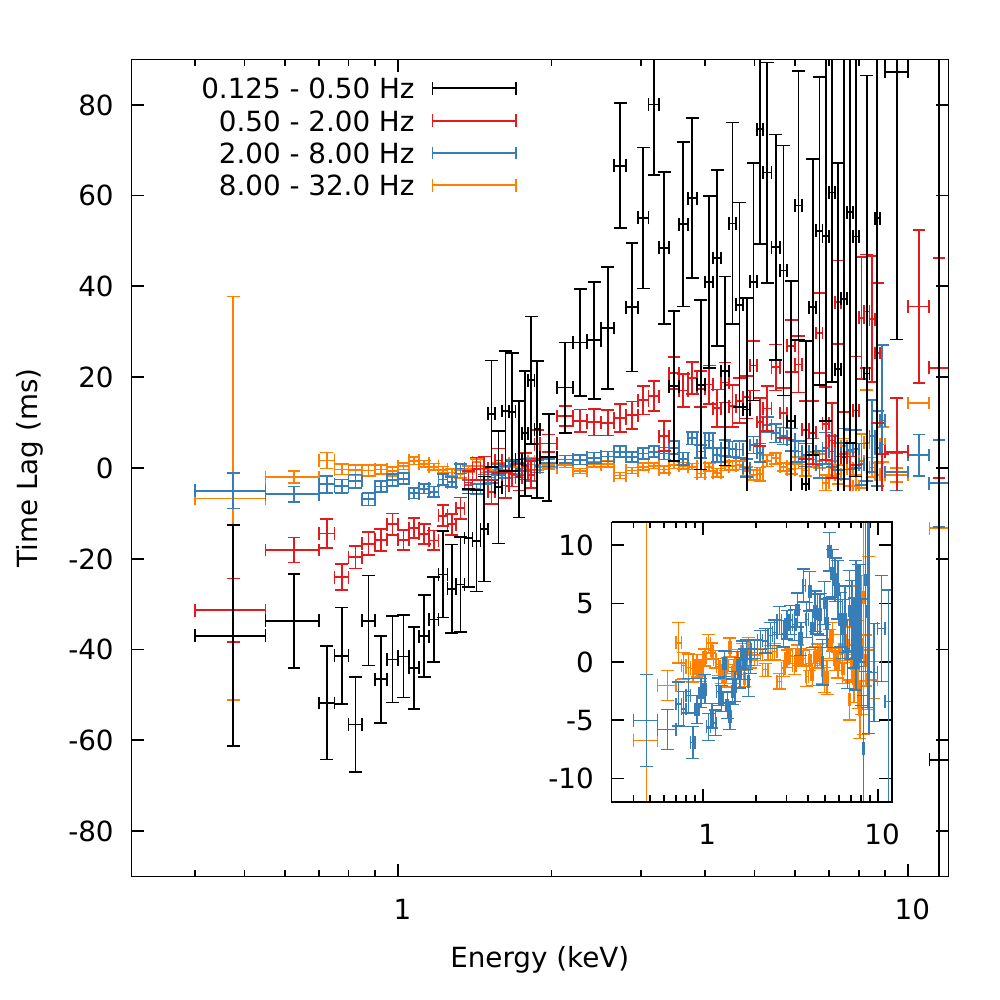}
    \caption{%
        Time lags {with respect to a broad reference band} in \aql 
        as a function of energy for four frequency
        intervals. The inset shows the two highest frequency intervals 
        over a smaller vertical range.
    }
    \label{fig:lags}
\end{figure}

    The time lags (Figure \ref{fig:lags}) consistently show that
    the soft emission leads the hard component. Additionally, we find
    that the size of the time lag decreases as a function of
    frequency. The slowest variations show a time difference between 
    $0.5$ keV and $10$ keV on the order of 100 ms, whereas the fastest 
    variability arrives nearly simultaneously. 

\begin{figure}[t]
    \centering
    \includegraphics[width=\linewidth]{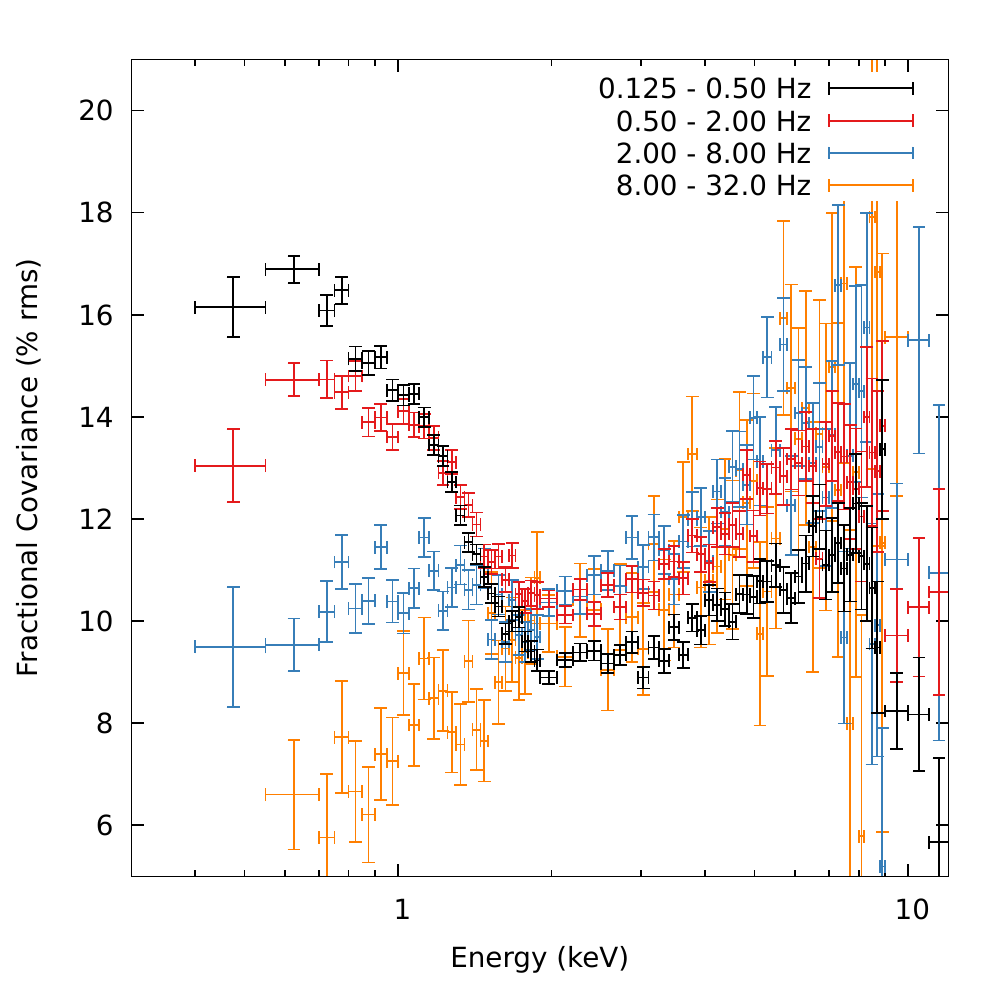}
    \caption{%
        Fractional covariance as a function of energy for four frequency
        intervals. 
    }
    \label{fig:frcov}
\end{figure}

    The covariance spectra, expressed as a fraction of the in-band
    count rate, are shown in Figure \ref{fig:frcov}. Consistent with
    the power spectral amplitudes, we find that the two
    lowest frequency bands show a dominant peak at 0.5 keV and
    turnover at 2 keV. In the 2--8 Hz band this soft component is 
    significantly weaker, and in the highest frequency band the
    turnover at 2 keV is no longer observed. Instead the two highest frequency 
    bands show a fractional variability that increases approximately linearly 
    from 0.5 to 10 keV.

    The $0.125-2$ Hz fractional amplitude of the covariance peaks at 
    17\% rms at 0.5 keV. This is slightly lower than the $25\%$ 
    peak fractional amplitude measured for the break component at $0.5$ keV as
    derived from the power spectrum. 
    This disparity is due to the different methods used to compute the
    amplitude. The covariance comes from integrating over a fixed frequency
    range, whereas the power spectrum amplitude comes from integrating over
    a Lorentzian profile. The latter captures a broader range of frequencies,
    and yields a higher integrated power. 

    In order to determine the association of the variability terms
    with the spectral components, we normalized the covariance spectra
    in terms of absolute rms amplitude and folded them with version 0.06 of the
    \nicer instrument response. We then fit the spectra in \textsc{xspec}
    v12.9.1.  
    
    We first fit the covariance spectrum of the $8-32$ Hz frequency band, as
    that has the simplest spectral shape. We model the Galactic absorption
    using the \texttt{tbabs} model using the abundances of \citet{Wilms2000}. 
    At a reduced $\chi^2$ of 1.04 (41 degrees of freedom, dof) the spectrum 
    is well described by an absorbed \texttt{powerlaw} model with photon index 
    $\Gamma = 1.2 \pm 0.1$. The absorption column is $N_{\rm H} = 6 \pm 1 \E{21}
    \mbox{ cm}^{-2}$, which is comparable to values reported from \textit{Chandra}
    and \xmm observations \citep{Campana2014}.
    
\begin{figure}[t]
    \centering
    \includegraphics[width=\linewidth]{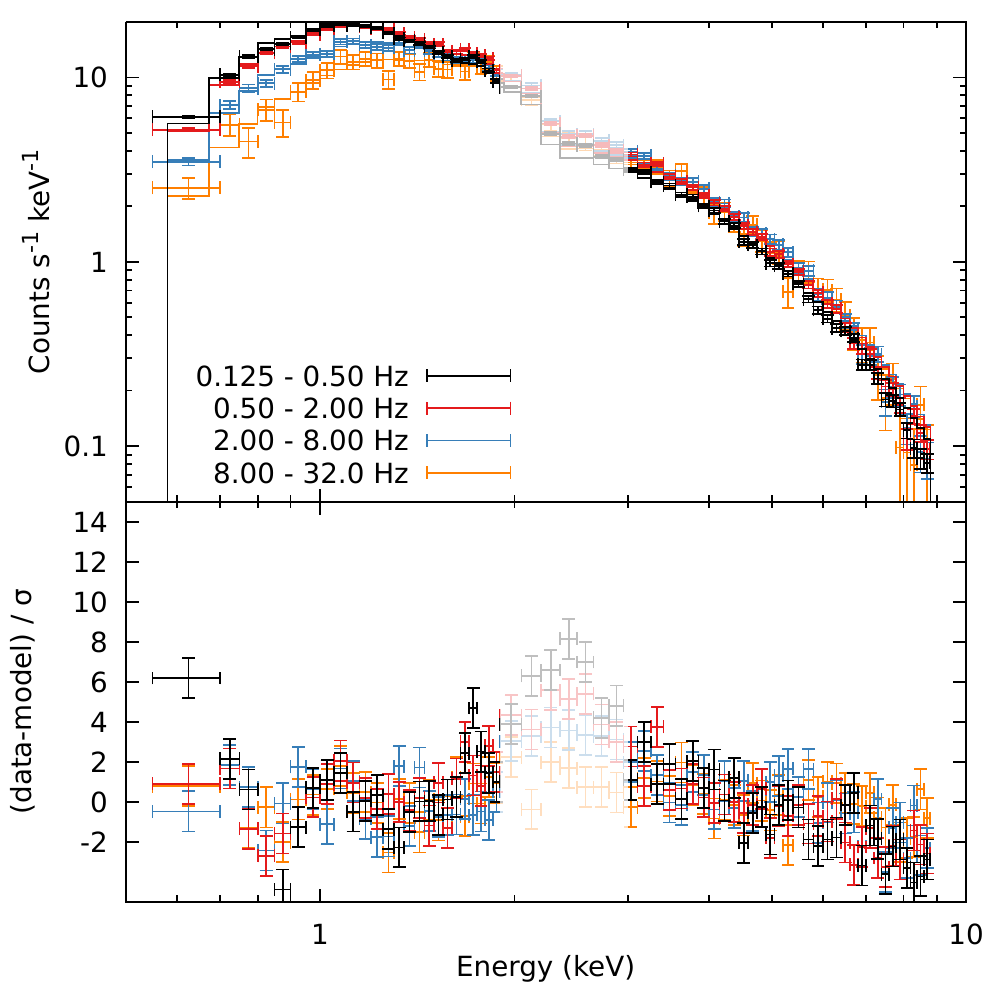}
    \caption{%
        Covariance spectra for four frequency intervals, with;
        top: the instrument-response folded spectra; bottom: 
        the best-fit residuals. Points in the $2-3$ keV range 
        were not included in the fit. 
    }
    \label{fig:covspec}
\end{figure}

    Having established a reasonable value for the absorption column, we proceed
    with performing a joint fit to all four covariance spectra.
    Compared to the initial $8-32$ Hz best-fit model, each of the other
    spectra show an unmodeled soft excess below 2 keV and a significant 
    residual near the known instrumental edge at 2.3 keV. Because efforts to 
    improve the instrument response matrix are still ongoing, we address 
    the instrumental feature by simply masking the spectral bins in the 
    $2-3$ keV range.
    The soft excess can be adequately described with either a single-temperature
    blackbody or a multi-temperature disk blackbody. 
    Adding an absorbed \texttt{diskbb} term to the spectral model we obtain 
    a temperature of $kT_{\rm in} = 0.24 \pm 0.01$ keV for a reduced 
    $\chi^2$ of 1.4 (135 dof; {Figure \ref{fig:covspec}}). 
    In this fit we kept $N_{\rm H}$ fixed and tied both disk temperature and the
    power-law photon index between each of the four covariance spectra. 
    The normalization of the power-law is approximately the same for each of
    the four spectra, whereas the normalization of the disk decreases with
    frequency. The complete set of fit parameters is shown in Table 
    \ref{tab:spec.fit}.

    Finally, we compare our model for the covariance spectra with the energy 
    spectrum of the time-averaged flux. Using all considered data we extract
    an energy spectrum in the 0.4--10 keV range, and extract a spectrum
    from \nicer observations of the \rxte background field 5 using identical
    filtering criteria. 
    After renormalization, the spectral model used for the covariance roughly 
    matches the shape of the time-averaged spectrum continuum, although
    significant residuals are present.
    The fit improves slightly by letting the time-averaged spectrum have 
    a different temperature and photon index, preferring $kT\simeq0.1$ keV
    and $\Gamma\simeq1.4$, respectively. Nonetheless, the fit remains only 
    approximate, and could also be described as a disk ($kT\simeq0.2$ keV),
    boundary layer ($kT\simeq0.8$ keV), and power-law ($\Gamma\simeq1.3$) akin
    to the model of \citet{Lin2007}.
	Neither model is statistically acceptable, and given the early state 
    of the instrument calibration it is difficult to be more rigorous 
    at this time (see \citealt{Ludlam2018}
    for a discussion of the calibration related uncertainties).
    
\begin{table*}[t]
    \newcommand{\mc}[1]{\multicolumn2c{#1}}
    \centering
	\caption{%
    	Covariance spectrum fit parameters
		\label{tab:spec.fit}	
	}
    \begin{tabular}{l l D D D D}
    \decimals
	\tableline
    Component & {Parameter} 
    &\mc{$0.125-0.5$ Hz}&\mc{$0.5-2$ Hz}&\mc{$2-8$ Hz}&\mc{$8-32$ Hz}\\
    \tableline	
\texttt{TBabs}    & $N_{\rm H}$ ($10^{22} \mbox{~cm}^{-2}$) 
										& 0.59    & 0.59    & 0.59    & 0.6(1) \\
\texttt{diskbb}   & $kT_{\rm in}$ (keV) & 0.24(1) & 0.24(1) & 0.24(1) & 0.24  \\
~                 & norm ($10^{ 2}$)    & 9.6(4)  & 7.4(3)  & 3.2(2)  & <0.7 \\
\texttt{powerlaw} & $\Gamma$            & 1.20(5) & 1.20(5) & 1.20(5) & 1.20(5) \\
~                 & norm ($10^{-2}$)    & 1.41(5) & 1.64(5) & 1.67(3) & 1.6(1) \\
	\tableline
	\end{tabular}
    \flushleft
    \tablecomments{
        Values in parentheses indicate $90\%$ confidence intervals, 
        and parameters without a quoted uncertainty were held fixed.
    }
\end{table*}

\section{Discussion}
    We analyzed the spectral-timing characteristics of \aql 
    in the hard state and find that the low-frequency band-limited
    noise exists in both the soft thermal emission and the hard
    power-law. Additionally, the variability of soft thermal
    emission leads the correlated modulation of the hard power-law.
    Hence, our results demonstrate that the thermal emission is 
    intrinsically variable and driving the band-limited noise 
    modulation of the power-law.
    
    The spectral-timing characteristics we observed in \aql are remarkably 
    similar to those seen in the hard state of the black hole binary GX 339--4
    \citep{Uttley2011}, and invite a similar interpretation.
    In the disk propagation model we can attribute the soft thermal 
    component of the energy spectrum to a cool accretion disk. This 
    disk has intrinsic variations with low frequencies that are
    observed directly at 0.5 keV, where the disk component dominates 
    the spectrum. Through mass accretion rate fluctuations, the variability
    can propagate radially down the accretion flow on a viscous timescale, 
    and modulate a central hot Comptonizing medium (e.g., a corona), which 
    gives rise to increasing variability above 2 keV. 
    The time-lag between the direct emission from the disk and the 
    reprocessed emission of the modulated Comptonizing region should 
    then scale with distance traveled, which is inversely related to 
    the considered frequency.
    
    We can compare GX 339--4 to \aql directly by considering the 
    three frequency bands in the $0.125-8$ Hz range, for which  
    covariance and lag measurements exist for both sources. For the black hole
    the three frequency bands have a $0.5$ to $10$ keV hard lag
    of $\sim150$ ms, $\sim20$ ms, and $\sim0$ ms, respectively.
    The hard lags measured in \aql are $\sim120$ ms, $\sim40$ ms, 
    and $\sim10$ ms. Hence, similar frequencies exhibit similar time lags.
    As a function of frequency the black hole time lags appear to
    have a steeper slope than the neutron star lags, albeit tentatively
    so.
    
    The fractional covariances are less straightforward to 
    compare, as spectral-timing studies of black holes do not 
    typically consider fractional amplitudes. However, we can 
    express the soft excess of our $0.125-0.5$ Hz covariance 
    spectrum as a ratio, by dividing the spectrum with the power-law 
    model component only. We then find the soft excess of \aql has a  
    ratio value of approximately $2.5$ at $0.5$ keV. This is again similar 
    to the GX 339--4 measurements of \citet{Uttley2011}.
    
    In contrast to GX 339--4, we find that the photon index of the covariance
    spectrum in \aql is comparatively low.  However, \aql has shown a brief
    epoch of coherent pulsations \citep{Casella2008} and spectral studies of
    its soft state suggest a truncated accretion disk \citep{King2016,
    Ludlam2017}. This suggests that the accretion flow in \aql may be
    interacting with a dynamically relevant magnetosphere, which could increase
    the electron temperature associated with the Comptonizing region, and
    possibly cause the flatter power-law.  We warn, however, that the photon
    index we report should be considered with some caution. A power-law model
    to describe the spectrum is almost certainly too simplistic, and is used
    only because our covariance spectra could not distinguish between more
    physically motivated models. A more sensitive approach would be to obtain
    the absorption column $N_{\rm H}$ and photon index from the time-averaged
    flux. 
    Given the early state of the instrument calibration, however,
    such a detailed study of the time-averaged flux cannot be robustly
    performed at this time.
    
    It is somewhat surprising that in the presence of a neutron star, 
    the hard state spectral-timing properties of \aql are so similar 
    to that of GX~$339-4$. 
    This may suggest that the emission associated with 
    the neutron star is not well correlated with the 
    band-limited noise, or that the spectral-timing signature 
    of that emission is very similar to that of the disk 
    propagation model.
    We note, for instance, that detailed modeling of a boundary 
    layer surrounded by a Comptonizing medium has been done 
    in the context of Z-sources. 
    Such work demonstrated that if the electron scattering optical 
    depth of the Comptonizing medium is modulated periodically, the
    emergent X-ray spectrum has similar characteristics to what
    we observe: time-lags between the emission above and below
    a pivot energy ($E_P$), and a minimum in the fractional
    variability at $E_P$ \citep{Miller1992, Lee1998}.
    We venture that a stochastic modulation of the
    optical depth may produce similar spectral-timing
    characteristics as seen in those calculations. The pivot
    energy would then be set by the shape of the Comptonized
    spectrum, and the power spectrum of the stochastic
    modulation. The slow modulation would still be driven by
    mass accretion rate fluctuation in the disk. Higher
    frequency fluctuations would then originate predominantly
    within the corona, and hence show increasing fractional
    covariance at high energies without generating significant
    time-lags. This picture does not fully address the
    similarities with black hole systems, but is at least
    qualitatively consistent with the spectral-timing properties
    we observe in \aql. We therefore suggest it may be
    worthwhile to explore whether such a model could
    reproduce the pivot energy and contribute to spectral-timing 
    properties we measured. 
    
    In summary, we have established that band-limited noise in the
    hard state of \aql is driven by the soft thermal component. 
    The striking similarities of our results with the black hole 
    GX~339--4 suggest that this soft component is most likely the 
    cool accretion disk, and that the spectral-timing features are 
    due to propagating mass accretion rate fluctuations. However,
    we note that the complex interactions of the accretion flow with 
    the neutron star surface and magnetosphere will require more
    detailed modeling to explain the similarity of observed 
    spectral-timing features. 
    We suggest that further spectral-timing studies of neutron stars, 
    enabled by \nicer, will provide fertile ground for these efforts.

\acknowledgments
This work was supported by NASA through the \nicer mission and the
Astrophysics Explorers Program, and made use of data and software 
provided by the High Energy Astrophysics Science Archive Research Center 
(HEASARC).    
PB was supported by an NPP fellowship at NASA Goddard Space Flight Center.  
EMC gratefully acknowledges support from the National Science Foundation 
through CAREER award number AST-1351222

\facilities{ADS, HEASARC, NICER}

\bibliographystyle{fancyapj}

\end{document}